\documentstyle[epsf, rotate]{l-aa}

\def\RXTE{{\it RXTE}}
\def\CGRO{{\it CGRO}}
\def\chisq{$\chi^2$}
\def\chisqnu{$\chi^2_\nu$}
\def\lesssim{\lower4pt \hbox{$\buildrel < \over \sim$}}
\def\gtrsim{\lower4pt \hbox{$\buildrel > \over \sim$}}

\begin{document}

\title{Radiation Feedback in Hot Accretion-Disk Corona Models and Application
to GX~339-4}
\author{M.  B\"ottcher, E. P. Liang, \and I. A. Smith}
\institute{Rice University, Space Physics and Astronomy Department, MS 108,
6100 S. Main Street, Houston, TX 77005 -- 1892, USA}
\date{Received ; accepted}
\offprints{M. B\"ottcher}

\maketitle
\markboth{M. B\"ottcher et al: Radiation feedback in hot accretion-disk
corona models}{M. B\"ottcher et al: Radiation feedback in hot accretion-disk
corona models}

\begin{abstract}
We present a detailed study of the observable effects of photoionization
and Comptonization of line and continuum radiation from a cold accretion
disk with a thin, warm, photoionized transition layer in the framework 
of self-consistent accretion-disk corona models for Galactic black-hole 
candidates. We use an iterative method to treat the non-linear 
radiation feedback between the transition layer and the hot corona 
numerically using a Monte-Carlo Comptonization code in combination 
with the photoionization and line transfer code XSTAR. The subset of the
parameter space allowed in self-consistent accretion-disk corona systems
on energetic grounds is explored, checking for the emergence of emission 
lines and/or absorption edges, with special emphasis on the spectral
range below a few keV, where such features might become observable
with the advent of the AXAF satellite, and investigating the
significance of the Compton reflection hump. Comparing our model 
calculations to the broadband spectrum of GX~339-4, we find good 
agreement with the observed spectral features, in particular the 
absence of a Compton reflection component. We discuss how the future 
detection of the predicted features at lower energies can be used 
to constrain parameters.

\end{abstract}

\section{Introduction}

The hard X-ray and soft $\gamma$-ray spectra from Galactic black-hole
candidates (GBHCs) in their low states can generally be well described 
by a power-law of energy index $\alpha \sim 0.4$ --- 0.9 with an 
exponential cut-off at $\sim$~100 -- 200~keV (e. g., Grove et al. 1998). 
In contrast to many Seyfert galaxies, which often show evidence for a 
pronounced bump around $\sim 20$ --- 100~keV and a strong Fe~K$\alpha$
emission line at 6.4~keV (e. g., Madjejski et al. 1995, 
Magdziarz et al., 1997), in Galactic black-hole 
candidates such features are often very weak or absent. In AGNs, they 
can be attributed to Compton reflection and iron fluorescence 
emission from a cold medium, either in a cold, optically thick 
outer accretion disk or in surrounding broad line regions (Lightman
\& White 1988, Matt et al. 1996; for a review see, e. g., Mushotzky 
et al. 1993). 

One of the popular models to explain the high-energy spectra of 
GBHCs consists of a cool, optically thick accretion disk surrounded 
by a corona of hot, optically thin plasma (Liang \& Price 1977,
Bisnovatyi-Kogan \& Blinnikov 1977, Liang \& Thompson 1979, 
Haardt \& Maraschi 1991, 1993). 
The detailed radiative transfer in a self-consistent two-phase pair 
corona model has been solved by Poutanen \& Svensson (1996). 
Based on energy balance arguments, Dove et al. (1997) 
determined the parameter space allowed for self-consistent accretion
disk corona (ADC) models, implying tight constraints on the Thomson 
depth $\tau_c$ and the temperature $T_c$ of the corona, which seem
to rule out a homogeneous slab geometry for the corona because the 
emerging spectra generally have energy spectral indices $\alpha 
\gtrsim 1$, inconsistent with the observed GBHC spectra. 

This motivated the patchy-corona model (Haardt, Maraschi \& 
Ghisellini 1994, Stern et al. 1995), in which the constraints 
on the Compton $y$ parameter of the active regions are relaxed 
in the sense that these regions can be significantly hotter 
than a uniform slab corona, leading to spectral indices $0.5 
\lesssim \alpha \lesssim 1$, depending on the local dissipation
compactness in the active regions (Stern et al. 1995).

In previous analyses of ADC models for GBHCs the cold disk was
generally a priori assumed to be only weakly ionized, leading 
to the typical Compton reflection bump, depleted by photoelectric
absorption, and an iron K$\alpha$ 
fluorescence line at 6.4~keV. However, independent studies of 
photoionization and fluorescence line emission (Matt et al. 
1993, $\dot{\rm Z}$ycki et al. 1994) 
from photoionized media have shown that the temperature
and ionization structure of the cold disk very strongly 
influences the shape of the resulting reflection spectrum,
regarding the continuum as well as line emission. These 
studies have also clearly demonstrated that the radiation
spectrum emitted by a photoionized disk strongly differs 
from the thermal blackbody spectrum usually assumed as the 
seed photon spectrum for Comptonization in accretion-disk
corona models. Given the moderate Thomson depth and
Compton $y$ parameter of the corona, $y = 4 \, (kT_c / 
m_e c^2) \, \tau_c \lesssim 1$ for the uniform corona and 
$y \sim 2$ for the patchy corona case, this has significant 
consequences for the emerging Comptonized spectrum,
particularly at low photon energies. With the advent
of the AXAF satellite, line features and absorption
edges in the spectra of GBHCs at $\lesssim$ a few keV 
might become observeable in spite of relatively strong 
interstellar photoelectric absorption by neutral material.

In this work, we adopt a method similar to the one used by
$\dot{\rm Z}$ycki et al. (1994), combining the XSTAR
photoionization and line transfer code (Kallman \& Krolik 1998,
Kallman \& McCray 1982) with our Monte Carlo Comptonization code 
in order to treat photoionization, line transfer, photoelectric
absorption and Compton reflection off the disk self-consistently. 
Using an iterative scheme, we investigate in detail the radiation 
feedback between the cold disk and the hot corona, accounting for
the effects of Comptonization of the cold disk emission in the 
hot corona and re-reflection of the Comptonized disk spectrum 
back onto the disk. 

In Section 2, we describe the model assumptions and the numerical
scheme used to treat the radiation feedback between the disk and
the corona. Results of our numerical simulations are presented
and discussed in Section 3. In Section 4, we demonstrate the
applicability of our model calculations to the observed X-ray /
soft $\gamma$-ray spectrum of GX~339-4 and make some predictions
for future observations of this object. We summarize in Section 5.

\section{Description of the model}

For our numerical simulations of the radiation feedback between
the cold accretion disk and the hot corona, we assume a plane-parallel 
geometry for either the entire corona (homogeneous corona) or for
individual active regions (patchy corona) located above the cold
disk. In the following, the subscript 'd' denotes quantities of
the cold disk, while the subscript 'c' refers to quantities of the 
corona. For the first step of our simulations, a cold disk temperature 
$T_d \sim 10^6$~K is used. The particle density in the cold disk 
is in the range $n_d \sim 10^{16}$ --- $10^{19}$~cm$^{-3}$. A
fraction $f_c$ of the cold disk is sandwiched by a hot 
corona of Thomson depth $\tau_c \sim 0.1$ --- 2 and 
temperature $kT_c \sim 30$ --- $150$~keV.

Given the radiation temperature of the cold disk as 
mentioned above, Dove et al. (1997) found strict upper 
limits for the temperature $T_c$ of a uniform corona, 
covering the entire disk, as a function of the Thomson depth. 
Stern et al. (1995) determined the allowed
parameter space in the case of a patchy corona. The corona 
parameters we will adopt in our numerical simulations are 
chosen to be consistent with these constraints. 

The details of the properties of the hot corona such as its 
density and temperature structure have a major influence 
only on the high-energy tail of the radiation spectrum.
Since in this paper we are mainly interested in the emergence of
line features and absorption edges at lower photon energies,
we assume for simplicity that the active regions of the corona 
have a uniform density and temperature in accordance with the 
general considerations by Dove et al. (1997) and Stern et al. 
(1995), and defer a detailed treatment of thermal and pair 
balance in the corona, taking into account the spectral shape 
of the accretion disk radiation and the physical shape of
the active regions, to future work. Here, we are fixing $T_c$
and $\tau_c$ without normalizing the total energy flux.

\begin{figure}
\rotate[r]{
\epsfysize=6cm
\epsffile[20 0 580 500]{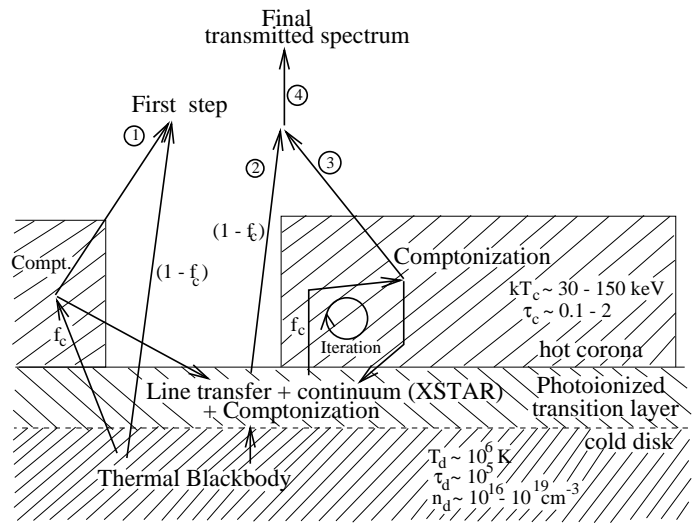}
}
\caption[]{Sketch of the model geometry and the iterative, numerical 
scheme used to treat the non-linear cold disk --- hot corona feedback}
\end{figure}

The temperature and ionization structure as well as the continuum
and line emissivities in the cold disk are calculated using the 
XSTAR code. However, since the distortion of the continuum spectrum 
by Compton scattering is neglected in XSTAR, we are using 
our own Monte Carlo Comptonization code in order to treat
Compton scattering in the photoionized surface layer of the 
cold disk self-consistently. The continuum radiation from the 
near-equatorial parts of the disk, at large Thomson depths from 
the disk surface, is represented by a thermal blackbody spectrum 
illuminating the transition layer from below. 

The geometry and the iterative numerical scheme we use are sketched
in Fig. 1. In the first step, we assume that the disk only emits a
thermal blackbody according to our first-guess temperature of
$T_d = 10^6$~K. The transition layer is defined as the portion
of the disk which is heated by the photoionizing continuum to
temperatures exceeding $T_d$. Its Thomson depth is typically 
$\sim 10$. We use an improved version of the Monte Carlo
Comptonization code developed by Canfield et al. (1987)
in order to simulate the processing of the disk radiation through the
corona, allowing for a geometry-dependent covering fraction $f_c$,
parametrizing the patchyness of the corona ($f_c = 1$ corresponds
to the uniform slab geometry). A fraction $(1 - f_c)$ of the disk 
spectrum emerges directly. 

We sample the transmitted spectrum as well as the spectrum of 
radiation reprocessed onto the cold disk. The reprocessed spectrum
from the corona is used in XSTAR to calculate the photoionization 
and temperature structure of the cold disk and the line and
continuum emissivities in the disk that are used to calculate
the radiation transfer due to Comptonization in the photoionized
transition layer, using our Monte-Carlo code. We assume the entire
disk to be uniformly illuminated by the coronal radiation, even
in the case of a patchy corona. Standard solar metal abundances 
in the cold disk are assumed.

In the second step, a fraction $f_c$ of the new disk spectrum is 
processed through the corona, and again the transmitted spectrum 
as well as the spectrum reprocessed onto the cold disk are sampled. 
The reprocessed disk spectrum is then used to re-calculate the disk 
ionization and temperature structure and continuum and line 
emissivities in the disk transition layer, which are used to
calculate a refined disk spectrum. This scheme is repeated iteratively 
until it converges to a stable output spectrum, if it does
converge. This is not always the case if the system results 
in a highly ionized, very hot accretion disk for which our
treatment is not valid. This issue will be addressed in more 
detail in the next section.

\begin{figure*}
\rotate[r]{
\epsfysize=12cm
\epsffile[20 0 580 500]{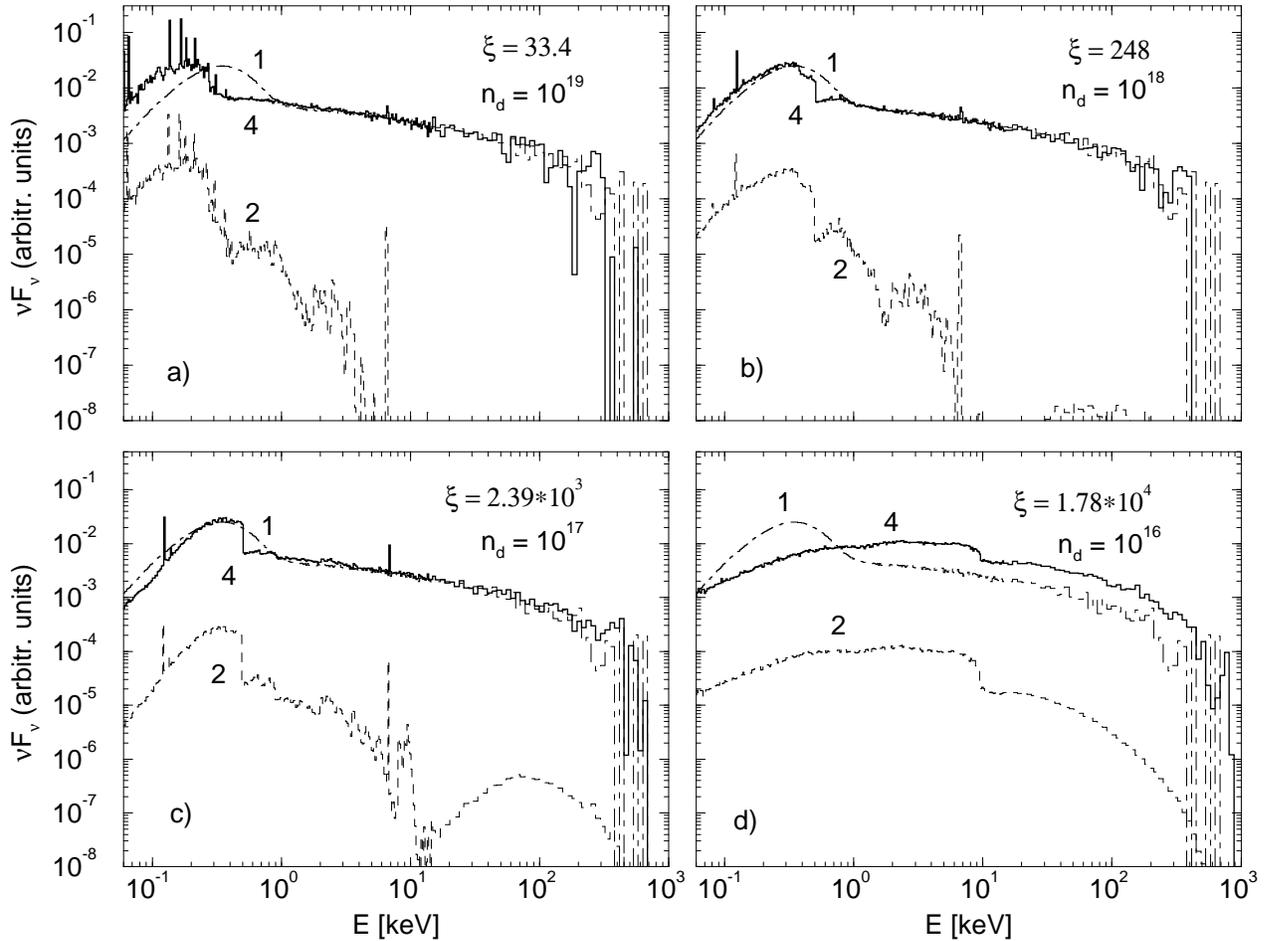}
}
\caption[]{The influence of variations of the ionization parameter
(given by variations of the density in the disk) on the Compton
reflection and line spectrum from the cold disk and the resulting
total spectrum. In all panels, we used $\tau_c = 0.2$, $kT_c = 140$~keV, 
and $f_c = 99$~\%. $n_d$ is in units of cm$^{-3}$, and $\xi$ in units 
of erg~cm~s$^{-1}$. The dashed curves (labeled `2') show the un-Comptonized 
part $(1 - f_c)$ of the cold disk spectra; the solid curves (labeled `4')
show the total emerging spectra. For comparison, the Comptonized thermal 
blackbody spectra (first guess for the disk spectrum) are shown by the
dot-dashed curves (labeled `1'). The labels correspond to the encircled 
numbers in Fig. 1. (The Comptonized disk spectrum component '3' is
identical to the total spectrum '4' in the cases shown here)}
\end{figure*}

The proper energy balance between irradiation of the cold disk 
by the corona and the emission from the disk is checked a posteriori, 
allowing for a fraction $f_v \lesssim 0.5$ of viscous heating with 
respect to the total heating rate (viscous heating + irradiation by 
the corona) within the cold disk.

The total luminosity resulting from the cold disk radiation 
and its Comptonization in the corona can be estimated by 
$L \sim A_d \, \sigma T_d^4 \, (1 + 4 \, f_c \, \Theta_c 
\tau_c)$, where $A_d$ is the effective disk area, $\sigma$ 
is the Stefan-Boltzman constant, and $\Theta_c = kT_c / 
(m_e c^2)$. Parameterizing the effective disk area in 
terms of Schwarzschild radii, $A_d = \pi \, (100 \, r_2 
R_S)^2$, this yields $L \sim 3 \cdot 10^{35} (M / M_{\odot})^2 
\, r_2^2 \, T_{d,6}^4$~erg~s$^{-1}$, where $T_{d,6} = T_d / 
(10^6 \, {\rm K})$ and $r_2$ is the effective radius of the 
disk in units of $100 \, R_S$. The average density of the 
disk can be related to the viscosity parameter $\alpha$ and
the accretion rate $\dot M = \dot m \, \dot M_{cr}$ with
$\dot M_{cr} = 3 \cdot 10^{-8} \, (M_{\odot} / yr) \, (M / 
M_{\odot})$ by $n_d \approx 4.3 \cdot 10^{17} \, \alpha^{-1} 
\, \dot m^{-2} \, (M / M_{\odot})^{-1} \, (\overline R / 6 
\, R_s)^{3/2}$ (Shakura \& Sunyaev 1973), where $\overline R$ 
is a typical radius characterizing the efficiently radiating 
portion of the disk.

\section{Numerical results and discussion}

Using the radiation-feedback code described in the previous section,
we explored a wide range of allowed (self-consistent) ADC model 
parameters. The restrictions for the corona parameters were found 
to be rather insensitive to the exact value of $T_d$ (Dove et al.
1997, Stern et al. 1995). They can be characterized by the fact 
that for a uniform corona the Compton $y$ parameter is restricted 
to $y_{\rm uniform} \lesssim 1$ (Liang 1979), while in a strongly 
patchy corona $y_{\rm patchy} \sim 2$.

\begin{figure*}
\rotate[r]{
\epsfysize=12cm
\epsffile[20 0 580 500]{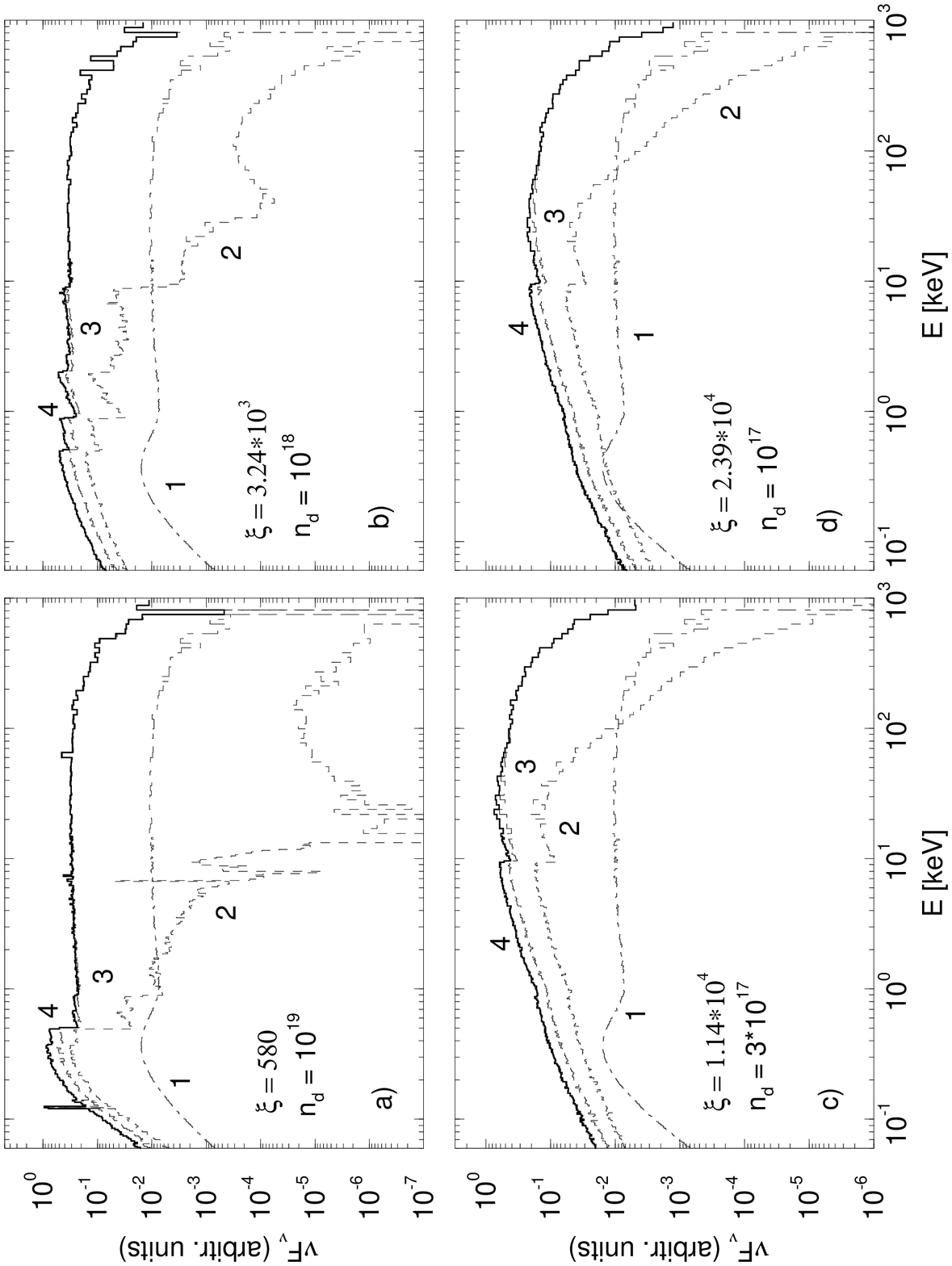}
}
\caption[]{The influence of a varying ionization parameter for 
a covering fraction of $f_c = 80$~\% and coronal parameters $kT_c 
= 100$~keV, $\tau_c = 0.5$. $n_e$ is in cm$^{-3}$, $\xi$ in
erg~cm~s$^{-1}$. The long-dashed curves (labeled `3') 
show the Comptonized disk spectrum; all labels correspond to the
encircled numbers in Fig. 1}
\end{figure*}

The ionization state of the matter in the cold disk is primarily
determined by the ionization parameter

\begin{equation}
\xi = {4\pi F_c \over n_d}
\end{equation}
where $F_c$ is the ionizing flux from the corona onto the disk. 
The effect of a varying ionization parameter, determined by a 
change in the assumed density of the transition layer of the 
cold disk, is shown in Fig. 2 where the ionization parameter 
is increased from $\xi = 33.3$~erg~cm~s$^{-1}$ (a) to $\xi = 
1.78 \cdot 10^4$~erg~cm~s$^{-1}$ (d). The coronal parameters were 
chosen to be $\tau_c = 0.2$ and $kT_c = 100$~keV, and the covering 
fraction is $f_c = 0.99$, basically corresponding to a homogeneous 
corona. It is important to note that $\xi$ is not a free parameter 
in these simulations, but is determined self-consistently through 
the illumination by the corona. When varying the ionization 
parameter, two important trends can be observed:

(1) In agreement with the results of Matt et al. (1993), 
the intensities of lines above $\sim 1$~keV show a maximum 
at a value of the ionization parameter around $\xi \sim 
2000$~erg~cm~s$^{-1}$. Near this maximum, strong iron K 
and L lines as well as strong absorption edges are seen in
the disk spectrum. The K line complex at 6.7 -- 9~keV is almost
completely smeared out by Comptonization in the corona, while the
lines and absorption edges below $\sim 1$~keV are clearly seen in
the transmitted spectrum. For significantly higher values of the 
ionization parameter, $\xi \gtrsim 10^4$, the fraction of completely 
ionized metals increases and the recombination rate decreases as a 
consequence of a high electron temperature, leading to a weakening 
of the line fluxes. Note that in the very-high ionization case, the
dominant absorption edge is not the Fe~K edge, because iron is 
almost fully ionized, but the Ni~K edge at 8.3~keV. For 
$\xi \gtrsim 2 \cdot 10^4$~erg~cm~s$^{-1}$,emission lines 
become insignificant. For much lower values of ionization 
parameter, $\xi \ll 10^3$, photoionization becomes less 
important, and the luminosity in fluorescence lines decreases 
compared to the maximum at $\xi \approx 2000$~erg~cm~s$^{-1}$.

\begin{figure*}
\rotate[r]{
\epsfysize=12cm
\epsffile[20 0 580 500]{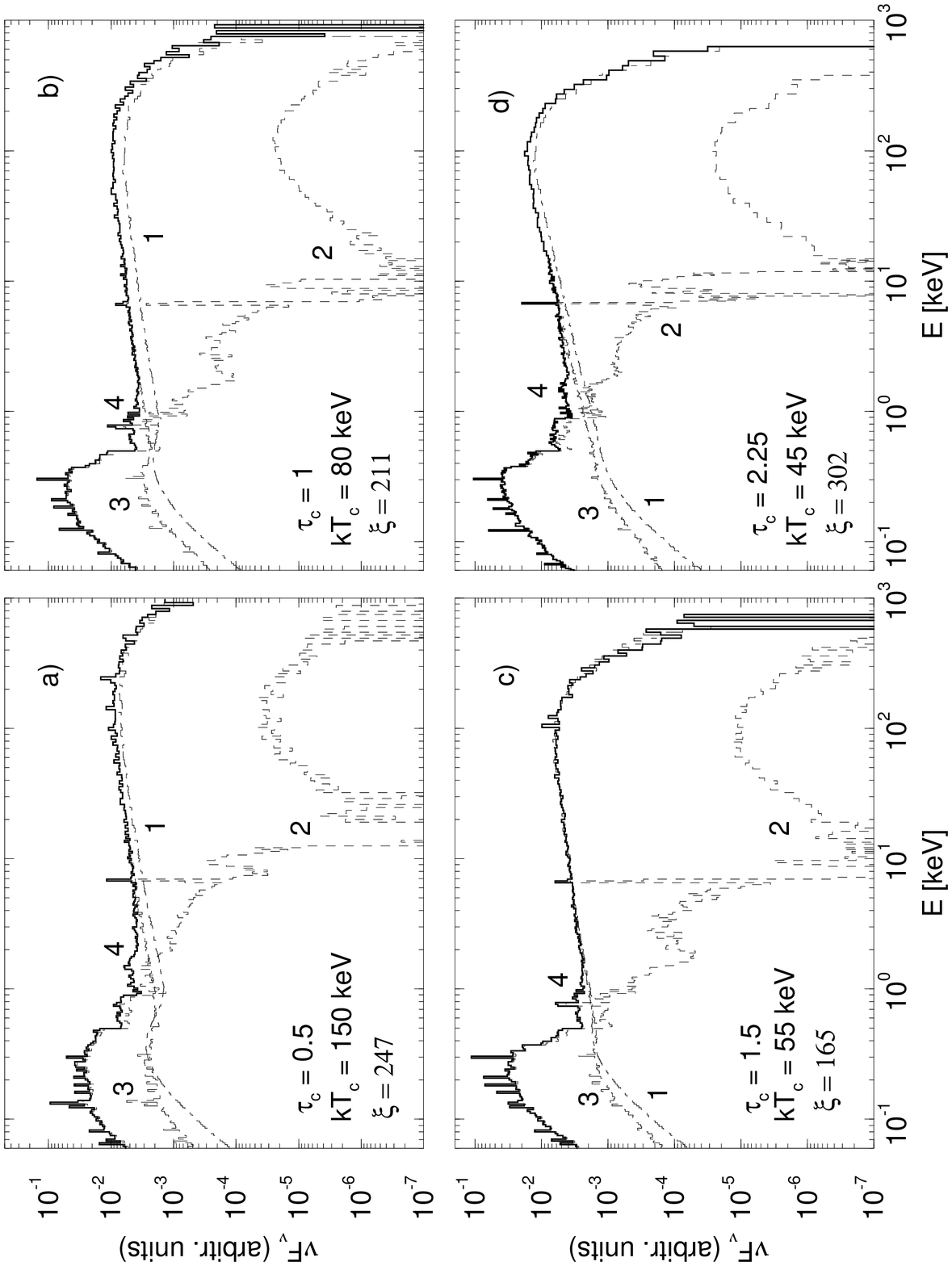}
}
\caption[]{The effect of varying the coronal parameters in the case
of a low covering fraction, $f_c = 15$~\%. $n_d = 10^{18}$~cm$^{-3}$.
$\xi$ is in erg~cm~s$^{-1}$. The curves are labeled according to the
encircled numbers in Fig. 1}
\end{figure*}

(2) It is remarkable that for moderately ionized disks, the Compton
reflection hump is very weak compared to the intrinsic disk
emission which is primarily due to bremsstrahlung and radiative 
recombination. This means that {\it most of the incident flux 
from the corona goes into heating of the disk surface layer and is 
not reflected}. As the ionization parameter increases, photoelectric 
absorption becomes less significant since an increasing fraction 
of the metals is highly or even completely ionized. Therefore, 
the Compton reflection hump becomes more luminous. In the limit 
of a very high $\xi$, the reflection spectrum does not produce the
typical bump around 20 --- 100~keV (depleted at lower energies
by photoelectric absorption), but it is itself a power law of
basically the same spectral index as the incident flux from the
corona (cf. $\dot{\rm Z}$ycki et al. 1994). 

It is obvious from Fig. 2 that in the case of a large covering 
fraction (homogeneous corona) and a moderate ionization parameter 
$\xi \ll 10^3$, corresponding to disk particle densities of 
$n \gg 10^{17}$~cm$^{-3}$, only a very weak iron K line feature 
around $\sim 7$~keV in the transmitted hard X-ray spectrum results,
although this line is quite prominent in the disk spectrum itself. 

Our results indicate that the Compton reflection component 
only results in a very weak flattening of the transmitted
Comptonization spectrum compared to the case of thermal 
Comptonization of a blackbody disk spectrum. This flattening 
is not sufficient to resolve the $\alpha < 1$ problem of the 
homogeneous corona models. For low ionization parameters, 
$\xi \ll 10^3$, the high-energy spectrum is virtually 
indistinguishable from the spectrum resulting from pure 
Comptonization of a thermal blackbody spectrum (i. e., 
curve 4 $\approx$ curve 1 in Fig. 2).

Fig. 3 illustrates the effect of a varying ionization parameter with
a lower covering fraction, $f = 80 \%$, corresponding to a slightly
patchy corona. For such a high covering fraction, the results do
not deviate very much from the case $f_c = 100$~\%. 

In the case of a very low covering fraction, $f_c \lesssim 20$~\%, 
enabling the choice of coronal parameters consistent with the 
observed hard power laws with $0.5 \lesssim \alpha \lesssim 1$ 
in the hard X-ray range, the transition to a highly ionized disk 
becomes very sensitively dependent on the covering fraction for 
disk densities $n_d \lesssim 10^{18}$~cm$^{-3}$. This is a result 
of the highly non-linear character of the radiation feedback 
discussed here. However, it should be pointed out that in the 
high-ionization case ($\xi \gtrsim 10^4$) a Thomson thick surface 
layer of the disk becomes very hot ($kT \gtrsim 10$~keV) and 
therefore itself emits a hard spectrum extending up to 
$\sim 100$~keV (see Fig. 2d, 3c,d). With this type of hardened
seed photon spectrum, the considerations regarding the energy 
equilibrium in the corona become invalid since these were 
based on a soft blackbody spectrum of temperature $\sim 10^6$~K. 
In some cases we even find that our numerical scheme does not 
converge in this situation, resulting in an accretion disk
approaching the coronal temperature and too high an ionization 
parameter to be accepted by the XSTAR code. A self-consistent 
treatment of this case therefore requires the detailed 
re-consideration of heating and cooling processes in the 
corona, using the exact disk spectrum, which we defer to 
future work. The extremely sharp transition between the 
medium-ionization and the high-ionization case might be 
an artifact due to the neglect of the detailed energy 
balance. The results presented here may therefore be 
regarded as reliable only in the case of moderate ionization 
of the disk ($\xi \lesssim 2000$~erg~cm~s$^{-1}$).

Varying the coronal temperature $T_c$ and Thomson depth $\tau_c$ 
within the allowed limits has only a very weak effect on the
resulting output spectrum, as long as the sharp transition to the
high-ionization case not reached. Some examples are illustrated 
in Fig. 4. With increasing Thomson depth and decreasing coronal 
temperature the line features emitted from the disk become more 
strongly smeared out in the transmitted spectrum. 

\begin{figure}
\rotate[r]{
\epsfysize=7cm
\epsffile[20 20 570 570]{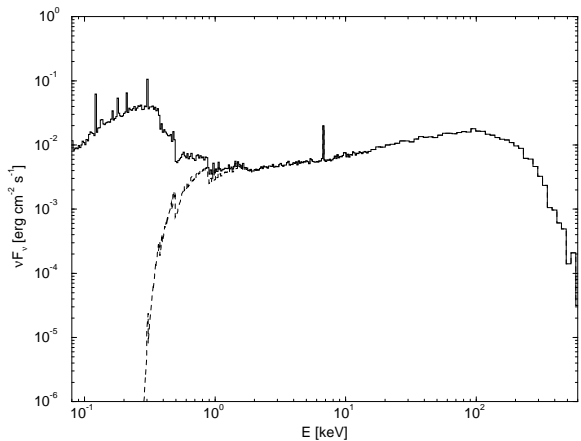}
}
\caption[]{The total disk corona spectrum of Fig. 4d (solid),
corrected for photoelectric absorption by neutral ISM or CSM, 
$N_H = 5 \cdot 10^{21}$~cm$^{-2}$ (dashed).}
\end{figure}

As mentioned above, the observed hard power-laws of the 
Comptonization continuum can be produced self-consistently
only in the patchy-corona model, invoking Thomson thick active 
regions. A second argument for the patchy corona are the observed
uncorrelated variability patterns of the hard power-law
component and the soft (cold disk) radiation component, which 
cannot be understood in the framework of a corona model with 
$f_c \sim 100$~\%.

Our results confirm that the exact spectrum of the cold disk 
has negligible influence on the hard power-law portion of the
X-ray spectrum. However, below a few keV, strong emission
lines from mildly to highly ionized metals may arise, depending
on the ionization parameter and the spectrum of the ionizing
continuum incident from the corona. However, at these energies
the X-ray flux from any source near the galactic plane will
be heavily absorbed by neutral ISM. To illustrate this, 
we plot in Fig. 5 the total AD corona spectrum of Fig. 4d, 
corrected for photoelectric absorption by a hydrogen column 
density of $N_H = 5 \cdot 10^{21}$~cm$^{-2}$, which is the 
estimate on $N_H$ used to model GX~339-4. It becomes obvious
that emission line features would be detectable down to 
$\gtrsim 0.3$~keV using observing instruments sensitive in
this energy range. 

\section{Application to GX~339-4}

In 1996, we performed a series of multiwavelength observations of the
Galactic black hole candidate GX 339--4 (Smith et al. 1998a,
Smith \& Liang 1998, Smith et al. 1998b).
GX 339--4 is unusual in that it is a persistent source, being detected 
by X-ray telescopes most of the time, but it also has nova-like flaring 
states. Our observations were made when the source was in a persistent
hard state (= soft X-ray low state).

As part of this campaign, we performed a pointed observation 1996 July 
9-23 using the Oriented Scintillation Spectrometer Experiment (OSSE) on 
the {\it Compton Gamma-Ray Observatory} (\CGRO). A pointed observation 
was also made using the {\it Rossi X-Ray Timing Explorer} (\RXTE) 1996 
July 26 (MJD 50290), just after the OSSE run ended. Full details of the
observations, the extraction of the spectra, and fitting of the spectra 
using simple phenomenological models is given in Smith et al. (1998a).

We generated the \RXTE\ spectrum using two of its instruments, the 
Proportional Counter Array (PCA), and the High Energy X-ray Timing 
Experiment (HEXTE). Our \RXTE\ observations showed that the emission 
was extremely variable, with the 2--5~keV band being slightly more 
variable than the 10--40~keV band (Smith \& Liang 1998). However, the 
hardness ratio showed no correlation with time during the observation, 
and so we averaged all the \RXTE\ data to generate a spectrum
that is representative of its shape throughout the run. Both a 
power law times exponential (PLE) and a Sunyaev-Titarchuk 
(ST; Sunyaev \& Titarchuk 1980) function fit the spectrum 
above $\sim 15$~keV. An additional soft component is required, 
as well as a broad emission feature centered on $\sim 6.4$ keV 
whose equivalent width is $\sim 600$~eV. No reflection component 
is required to fit the spectrum.

The OSSE spectrum was extracted by averaging over the whole two 
week observation. Since the hardness ratio did not change significantly 
over this time, this again gives a reliable measure of the spectral 
shape. Both a PLE and a ST model also fit our OSSE spectrum on its 
own.

We combined the \RXTE\ and OSSE data to generate the joint spectrum.
The results should be treated with care, because the two data sets are 
not quite simultaneous, and the source is highly variable. However, 
since the hardness ratios did not change dramatically in either 
observation, the resulting spectral shape is a good representative 
average of the source in 1996 July. We found that the PLE model easily 
fits the joint spectrum. However, the ST model has too much curvature 
to give an acceptable joint fit.

\begin{figure}
\rotate[r]{
\epsfysize=7cm
\epsffile[20 20 570 570]{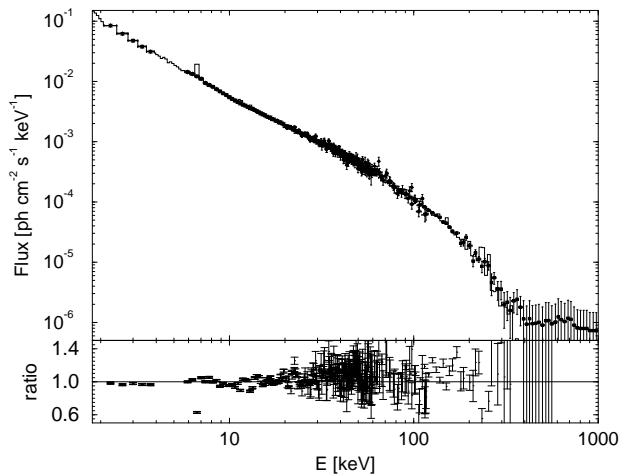}
}
\caption[]{Joint \RXTE--OSSE spectrum for 1996 July observations 
of GX 339--4. Note that the two data sets are not quite 
simultaneous, and the source is highly variable, so this figure should 
be used with care. The data have been unfolded using a PLE model with 
a power law for the soft component plus a Gaussian line and an edge.
The histogram shows a fit by eye using our detailed corona model. The
lower panel shows the ratio data/model}
\end{figure}

Figure 6 shows the unfolded data using a PLE fit (this data is
from Fig. 9 of Smith et al. 1998a). The PCA, HEXTE, 
and OSSE data have been normalized as described in Smith et al. 
(1998a). In the PLE model, the flux has the form 
$F(E) \propto E^{-\alpha} \exp(-E/kT)$, and the best fit values were
$\alpha = 1.22 \pm 0.01$ and $kT = 96.9$ keV (fixed by the OSSE data).
In addition to the PLE, the soft component is a power law with photon 
index $3.00 \pm 0.06$ and $N_H = 5 \times 10^{21}~{\rm cm}^{-2}$ (fixed
from the {\it EXOSAT} observations that more reliably measured the 
lower energy spectrum; Ilovaisky et al. 1986, M\'endez \& van der 
Klis 1997). On the basis of ROSAT observations, Predehl et al.
(1991) found $N_H = 6 \times 10^{21}$~cm$^{-2}$, which is consistent
with the value found by Zdziarski et al. (1998). However, the determination 
of $N_H$ on the basis of X-ray observations depends on the assumption 
on the spectral shape at low energies, which is uncertain. The exact
value of $N_H$ is not critical for the fitting shown here, but does
impact the future detectability of the features below a few keV.
To the phenomenological model a Gaussian line is added, with centroid 
$6.6 \pm 0.3$ keV and width $\sigma = 1.7 \pm 0.1$~keV, and an edge 
with threshold energy $7.18 \pm 0.08$~keV and absorption depth 
$0.09 \pm 0.02$ at the threshold is included. (The errors are 90~\% 
confidence regions for varying one parameter.) This fit has reduced 
\chisqnu = 1.025. The probability that a random set of data points 
would give a value of $\chi^2$ as large or larger than this is 
$Q = 0.36$.

Our detailed model is highly non-linear and cannot be written 
as a simple parametrization. We are therefore currently unable 
to directly fit the observations using the normal forward folding 
\chisq\ minimization techniques. Eventually we aim to build a 
library of spectra generated for ranges of parameters of interest, 
and then be able to perform the fitting from these. In this paper, 
we restrict ourselves to fitting by eye the unfolded spectrum 
obtained from the PLE best fit model to illustrate that our model 
gives realistic results.

The histogram in Figure 6 shows a realistic model calculation using 
our detailed corona model. The parameters used are $\tau_c = 2.2$,
$kT_c = 44$~keV, $f_c = 18$~\%, $n_d = 5 \cdot 10^{17}$~cm$^{-3}$
with the resulting ionization parameter $\xi = 516$~erg~cm~s$^{-1}$.
The continuum shape over the entire energy range of the
observation is very well described by our model calculation.
However, since we neglected the effects of line broadening due
to Keplerian rotation and turbulent motion in the disk, the
model Fe~K$\alpha$ line is narrower than observed. 
Although the line width is not accurately modeled by our 
calculation, Fig. 6 indicates that, in addition to the hard
power-law and the exponential cutoff, the soft power-law at low
energies as well as the iron edge, required by the phenomenological
fit to the data, are reasonably well reproduced by our model.

In agreement with the phenomenological fit, the reflection component 
resulting from our model calculation is negligible. Our simulations 
show that this is a natural consquence of the moderate
ionization of the surface layer of the disk. 

The absence of a strong reflection component is also in agreement 
with the analysis of observations of Cyg~X-1 by Dove et al. 
(1998) two weeks after a transition to 
its hard (low) state, using a broadband spectrum covering 
the entire 3 -- 200~keV energy range. These authors argue 
that the strong Compton reflection component usually seen in 
the low state of Cyg~X-1 might in part be an artifact due to 
the incomplete frequency coverage of previous observations of 
Cyg~X-1. However, if they restrict their analysis to the 
3 -- 30~keV and 100 -- 200~keV ranges, they find a best fit 
including a relatively strong Compton reflection component 
corresponding to a covering fraction $f = 0.35 \pm 0.02$ of 
a cold disk intercepting the hard power-law emitted by the 
corona. This would be in agreement, e. g., with the results 
of Gierli\'nski et al. (1997) who used Ginga + OSSE observations,
covering approximately the energy range mentioned above.

If the value of $N_H = 5 \cdot 10^{21}$~cm$^{-2}$ for the galactic 
hydrogen column density towards GX~339-4, which we assumed in
unfolding the observed spectrum, is realistic, our calculations
predict that future observations of this object in the energy
range $\sim 0.3$ -- 1~keV, which will become possible with the
advent of the AXAF mission, should reveal significant emission
lines and/or absorption edges from the photoionized disk which
will be extremely useful to further constrain the ionization
state of the disk surface and thereby the geometry. Rotational
and turbulent broadening might smear out some of the weaker
lines at low energies. However, the absorption edges are
expected to remain detectable in any case. The model 
calculation shown in Fig. 6 predicts the presence of strong
Fe~L and Ni~L edges at 0.7 and 0.8~keV and a strong O~L edge 
at 0.5~keV. These predictions, in turn, can be used to determine
the hydrogen column density towards the source more precisely.

\section{Summary and conclusions}

We have developed a method to treat self-consistently the radiation
feedback between a cold accretion disk and a homogeneous or patchy,
hot corona surrounding the disk. The detailed shape of the disk
radiation spectrum, strongly deviating from a blackbody spectrum,
has consequences especially for the observable X-ray spectrum below
a few keV. Above this energy, the Comptonization spectra are
virtually indistinguishable from thermal Comptonization of a
soft blackbody spectrum. 

Accounting for the exact thermal and ionization balance, 
using the XSTAR code, we find that due to the usually 
moderate ionization of the surface layer of 
the cold disk, the reflection component is in most cases
negligible. This is in accord with recent results that the
broadband X-ray and soft $\gamma$-ray spectra of GX~339-4
and Cyg~X-1 can be well fitted with a pure power-law +
exponential cutoff, without a Compton reflection component.

Our model calculations predict the existence of strong
emission lines and/or absorption edges at energies $\lesssim 1$~keV,
which are usually hard to observe because of strong absorption
by neutral hydrogen. However, with the advent of the AXAF
satellite with its unprecedented sensitivity at low X-ray
energies, it might become possible to test the predictions
of our calculations and use the results to determine the
ionization parameter of the disk surface layer and the covering
fraction of the corona with respect to the cold disk. Given that
the coronal temperature and Thomson depth can usually be well
constrained by the hard power-law index and the cut-off energy 
at hard X-ray / soft $\gamma$-ray energies, knowledge of the
disk surface ionization state and the covering fraction would
allow a detailed study of the structure of the accretion disk.

\acknowledgements{We thank T. Kallman for advice regarding the XSTAR
code. This work was supported by NASA grants NAG~5-4055, NAG~5-1547,
and NAG~5-3824 at Rice University.}

\end{document}